\documentclass{nature}

\usepackage{amssymb}
\usepackage{amsmath}
\usepackage{graphicx}
\usepackage{url}
\usepackage{xcolor}
\title{The optical afterglow of the short gamma-ray burst associated with GW170817}

\author{
J. D. Lyman$^{1,\star}$,
G. P. Lamb$^{2,3}$, 
A. J. Levan$^{1}$, 
I. Mandel$^{4, 5}$, 
N. R. Tanvir$^{3}$,
S. Kobayashi$^{2}$,
B. Gompertz$^{1}$, 
J. Hjorth$^{6}$, 
A. S. Fruchter$^{7}$, 
T. Kangas$^{7}$,
D. Steeghs$^{1}$,
I. A. Steele$^{2}$,
Z. Cano$^{8}$, 
C. Copperwheat$^{2}$,
P.A. Evans$^{3}$, 
J.P.U. Fynbo$^{9}$, 
C. Gall$^{6}$,
M. Im$^{10}$, 
L. Izzo$^{8}$, 
P. Jakobsson$^{11}$, 
B. Milvang-Jensen$^{9}$,
P. O'Brien$^{3}$, 
J.P. Osborne$^{3}$, 
E. Palazzi$^{12}$,
D.A. Perley$^{2}$, 
E. Pian$^{12}$, 
S. Rosswog$^{13}$,
A. Rowlinson$^{14}$,
S. Schulze$^{15}$, 
E.R. Stanway$^{1}$, 
P. Sutton$^{16}$, 
C.C. Th\"one$^{8}$, 
A. de Ugarte Postigo$^{8,6}$,
D.J. Watson$^{9}$,
K. Wiersema$^{1}$        \& 
R.A.M.J. Wijers$^{17}$ 
}

\begin{document}

\maketitle

\begin{affiliations}
 \item Department of Physics, University of Warwick, Coventry, CV4 7AL, UK
 \item Astrophysics Research Institute, LJMU, IC2, Liverpool Science Park, 146 Brownlow Hill, Liverpool L3 5RF, UK
 \item Department of Physics and Astronomy, University of Leicester, LE1 7RH, UK
 \item Birmingham Institute for Gravitational Wave Astronomy and School of Physics and Astronomy, University of Birmingham, Birmingham, B15 2TT, UK
 \item Monash Centre for Astrophysics, School of Physics and Astronomy, Monash University, Clayton, Victoria 3800, Australia
 \item Dark Cosmology Centre, Niels Bohr Institute, University of Copenhagen, Juliane Maries Vej 30, Copenhagen \O, 2100, Denmark
 \item Space Telescope Science Institute, 3700 San Martin Drive, Baltimore, MD 21218, USA
 \item Instituto de Astrof\'isica de Andaluc\'ia (IAA-CSIC), Glorieta de la Astronom\'ia, s/n, 18008, Granada, Spain
 \item The Cosmic Dawn Center, Juliane Maries Vej 30, DK-2100 Copenhagen O, Denmark
 \item Center for the Exploration of the Origin of the universe (CEOU), Seoul National University, Seoul, Korea; Astronomy Program, Department of Physics \& Astronomy, Seoul National University, Seoul, Korea
 \item Centre for Astrophysics and Cosmology, Science Institute, University of Iceland, Dunhagi 5, 107 Reykjav\'ik, Iceland
 \item INAF, Institute of Space Astrophysics and Cosmic Physics, Via Gobetti 101, I-40129 Bologna, Italy
 \item The Oskar Klein Centre, Department of Astronomy, AlbaNova, Stockholm University, SE-106 91 Stockholm, Sweden
 \item Anton Pannekoek Institute, University of Amsterdam, Postbus 94249, 1090 GE Amsterdam, The Netherlands
 \item Department of Particle Physics and Astrophysics, Weizmann Institute of Science, Rehovot 761000, Israel
\item School of Physics and Astronomy, Cardiff University, Cardiff, United Kingdom, CF24 3AA
\item Anton Pannekoek Institute for Astronomy, University of Amsterdam, Postbus 94249, NL-1090 GE Amsterdam, the Netherlands
\end{affiliations}

\begin{abstract}
The binary neutron star merger GW170817 was the first multi-messenger event observed in both gravitational and electromagnetic waves.\cite{lvc17a,lvc17b} The electromagnetic signal began $\sim$ 2 seconds post-merger with a weak, short burst of gamma-rays,\cite{lvc17grb} which was followed over the next hours and days by the ultraviolet, optical and near-infrared emission from a radioactively-powered kilonova.\cite{arcavi17,coulter17,evans17,pian17,smartt17,soares17,tanvir17,valenti17} Later, non-thermal rising X-ray and radio emission was observed.\cite{haggard17,hallinan17}
The low luminosity of the gamma-rays and the rising non-thermal flux from the source at late times could indicate that we are outside the opening angle of the beamed relativistic jet. Alternatively, the emission could be arising from a cocoon of material formed from the interaction between a jet and the merger ejecta.\cite{hallinan17,mooley17,ruan17} Here we present late-time optical detections and deep near-infrared limits on the emission from GW170817 at 110~days post-merger. 
Our new observations are at odds with expectations of 
late-time emission from kilonova models, being too bright and blue.\cite{kasen17, waxman17}
Instead, the emission arises from the interaction between the relativistic ejecta of GW170817 and the interstellar medium.
We show that this emission matches the expectations of a Gaussian structured relativistic jet, which would have launched a high luminosity short GRB to an aligned observer.
However, other jet structure or cocoon models can also match current data -- the future evolution of the afterglow will directly distinguish the origin of the emission.
\end{abstract}

\begin{figure}
\centering
\includegraphics[width=\linewidth,trim={1.5cm 1cm 0 0}]{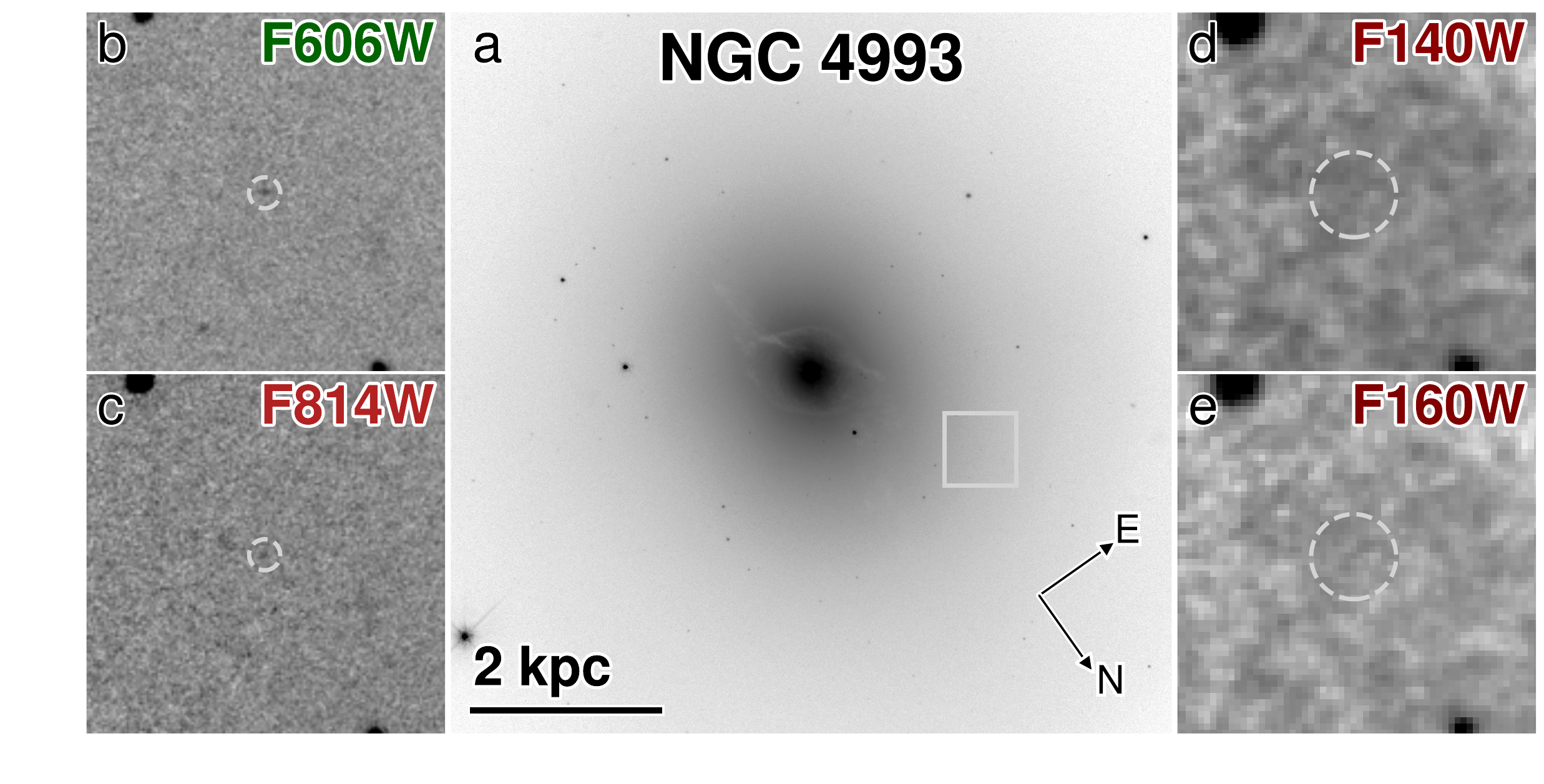}
\caption{{\bf \sffamily Late-time Hubble Space Telescope/Wide Field Camera 3 optical and near-infrared images of GW170817.} The host galaxy, NGC~4993 is shown in F606W ({\bf \sffamily a}). Orientation and linear scale are indicated, as well as the region covered by the zoom-in panels. The merger site in each filter, as indicated, are shown ({\bf \sffamily b-e}). The dashed circles (radii $\simeq$ 2.5 $\times$ FWHM) are centered on our determined location of GW170817 in each case. The zoom-in panels have had the underlying galaxy background light subtracted (see Methods). We find significant detections in F606W and F814W but only upper limits on the flux in F140W and F160W.
}
\label{img_fig}
\end{figure}

\begin{figure}
\centering
\includegraphics[width=\linewidth]{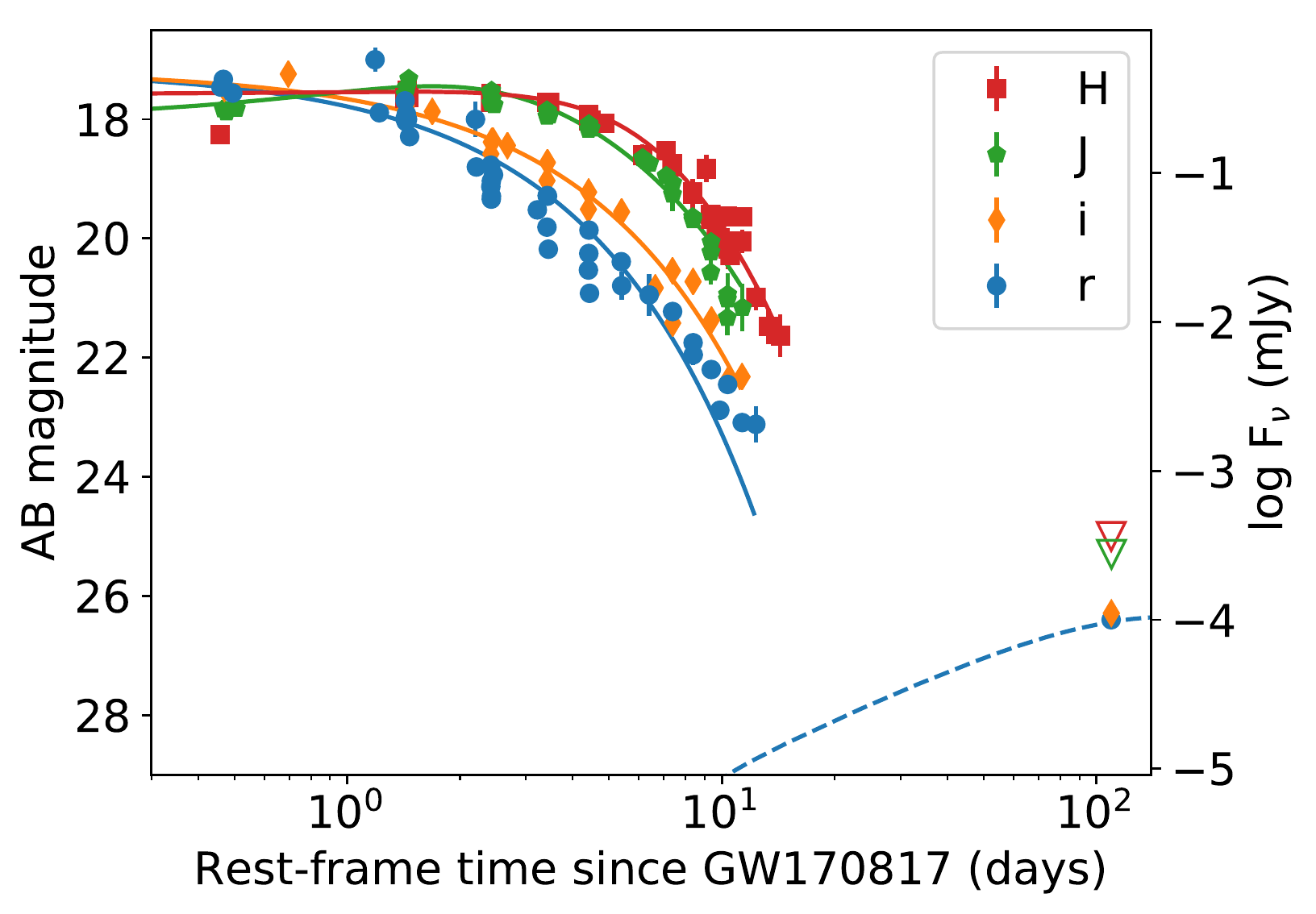}
\caption{{\bf \sffamily Optical and near-infrared light curve of GW170817.} Multiwavelength light curves of the counterpart of GW170817 including our new late time measurements. Ground-based $r$, $i$, J and H band filter light curves are shown by blue, orange, green and red markers respectively. Our late-time results are shown in the colour and marker of the most similar ground-based photometric filter; near-IR 3$\sigma$ limits are shown as open downward triangles. Earlier epochs are dominated by kilonova light, however the latest optical detections are much brighter and bluer than what is expected from kilonova models. The solid lines, in their respective filter colours, show the results of fitting an exponentially rising and decaying phenomenological function to the early light curves.\cite{gompertz17} The dashed line shows our structured jet afterglow in filter F606W ($\simeq$ $r$). Error bars indicate 1$\sigma$ uncertainties. Earlier photometry has been previously published\cite{arcavi17,coulter17,drout17,cowperthwaite17,kasliwal17,shappee17,smartt17,tanvir17,valenti17} and was taken from \url{https://kilonova.space}.\cite{guillochon17}
}
\label{lc_fig}
\end{figure}

For the Hubble Space Telescope ({\em HST}), the end of Sun constraint for GW170817 was on 6 December 2017 ($\sim 110$~rest-frame days post-merger), and we immediately obtained deep observations in the optical and infrared (see Table~\ref{obslog} and Methods for details of the observations and reduction). The new images were astrometrically aligned to our earlier epoch {\em HST} data in order to accurately locate the merger site and perform photometry (see Methods). Images of the merger site in each of our filters are shown in Figure~\ref{img_fig}. We detect emission at the location of the merger in the optical F606W and F814W filters (central wavelengths, $\lambda_\text{cen} \sim 589$, $802$\,nm, respectively). For the near-IR filters F140W and F160W ($\lambda_\text{cen} \sim 1392$, $1527$\,nm, respectively) we could not establish significant detections and so can place only upper limits on the transient flux at these wavelengths. Optical and near-infrared light curves for the counterpart to GW170817, including our recent observations, are shown in Figure~\ref{lc_fig}.

A detection in the optical or near-IR at such late times is not expected from the family of kilonova models currently in use. Indeed, most detailed studies stop at $\sim 30$\,days where predicted luminosities correspond to $\gtrsim 30$\,mag,\cite{rosswog17,kasen17} undetectable for even {\em HST}. Alternative models of kilonova emission with a slower decay of the light curves\cite{waxman17} 
would nevertheless predict redder emission than we observe. Initially blue, with $M_{r\rm,AB} - M_{H\rm,AB} \simeq 0.4$\,mag
at 1.5 days\cite{drout17}, GW170817 evolved to become very red, with 
$M_{\rm F606W,AB}-M_{\rm F160W,AB}= 2.8$\,mag 
at 11~days post-merger\cite{tanvir17}, consistent with optical line blanketing in the 
lanthanide-rich ejecta. Our late time detections and limits imply a much bluer colour at 110~days of $M_{\rm F606W,AB}-M_{\rm F160W,AB}\lesssim 1.5$\,mag. Such evolution from blue to red and back to blue is not expected from current kilonova models. 
We note that this colour is bluer than that of typical globular clusters, and the source fainter than the majority of them (Figure \ref{fig_gc}). We consider our detections as being due to the transient and not an underlying source, although longer term optical monitoring will rule conclusively (see Supplementary Information).

Instead, we consider our observations in light of the radio and X-ray detections of the non-thermal GRB afterglow radiation.  This synchrotron radiation is produced by relativistic electrons gyrating in a  magnetic field.  The  electrons in the interstellar medium around the merger may be accelerated to relativistic velocities by shocks arising from either a collimated, initially ultra-relativistic jet or a more isotropic, mildly relativistic `cocoon'.   Early X-ray non-detections made with the {\em Swift} satellite,\cite{evans17} followed by later detections of rising radio and X-ray flux,\cite{haggard17,hallinan17,ruan17} indicate that either the jet is being viewed off-axis, or that the jet is unable to punch through the dynamical ejecta from the merger and the cocoon model is correct.\cite{hallinan17,mooley17}
The continued gradual rise of the radio flux from 15 to more than 100 days is difficult to reconcile with a classical `top-hat' jet profile viewed off-axis, which would be expected to have a steeper rise (see Supplementary Information). A `top-hat' jet has a homogeneous energy distribution within the jet opening angle, which sharply drops outside the jet. In reality, jets are unlikely to show this morphology and several structured jet models have been proposed.\cite{kumargranot03,zhang04} The temporal behaviour of the afterglow light curve from a decelerating structured jet depends on the specific structure model and viewing angle.\cite{lamb17}

\begin{figure}
\centering
\includegraphics[width=\linewidth]{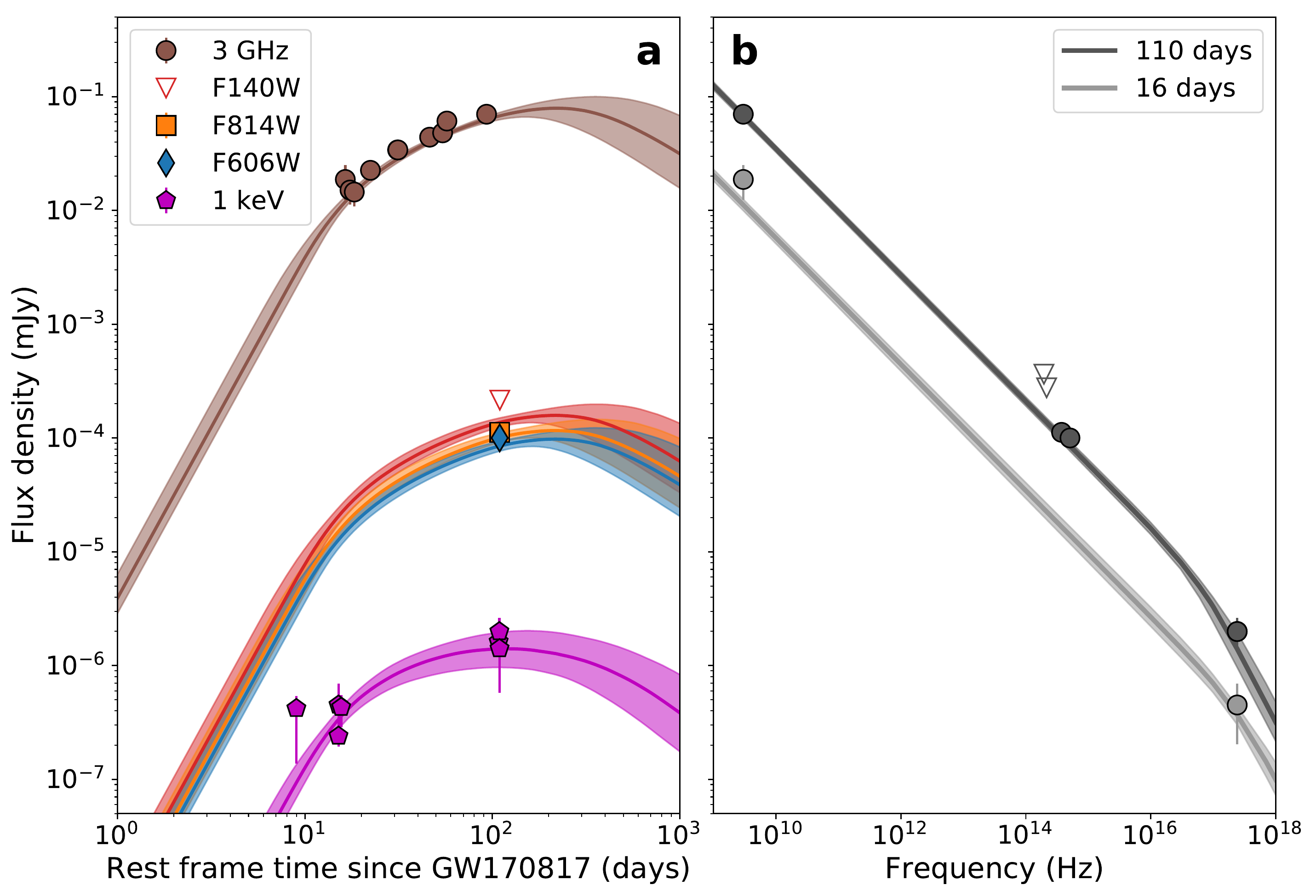}
\caption{{\bf \sffamily Light curve and spectra for Gaussian structured jet model.} The observed light curve and spectral energy distribution of the afterglow at radio, optical and X-ray wavelengths can be described by an off-axis structured jet. {\bf \sffamily a}: Model light curves from our MCMC implementation of the off-axis structured jet model. The thick lines denote the median light curve from 10000 MCMC samples, with the shaded regions indicating the 16 and 84th percentiles. Overlaid are 3~GHz radio\cite{hallinan17,mooley17} (brown) and 1~keV X-ray\cite{margutti17,haggard17,troja17,ruan17} (magenta) data on the appropriately coloured light curves. Our {\em HST} near-IR F140W flux limit is shown by an open triangle, with the F140W model light curve in red. Our optical F814W and F606W detections are denoted by orange square and blue diamond markers, respectively, again overlaid with the model light curve at these frequencies.
{\bf \sffamily b}: The model spectra at 16 days (light grey) and 110 days (dark grey) post-merger with the median and 16 and 84th percentiles indicated as for the light curves. These spectra are compared to almost contemporaneous radio\cite{hallinan17,mooley17} and X-ray\cite{haggard17,ruan17} data at each of the epochs. Our near-IR limits and optical detections are also indicated on the 110~day spectrum. Upper limits are shown as unfilled triangles. Uncertainties shown are 1$\sigma$ and are smaller than some markers.}
\label{modelfig}
\end{figure}

\begin{figure}
\centering
\includegraphics[width=\linewidth]{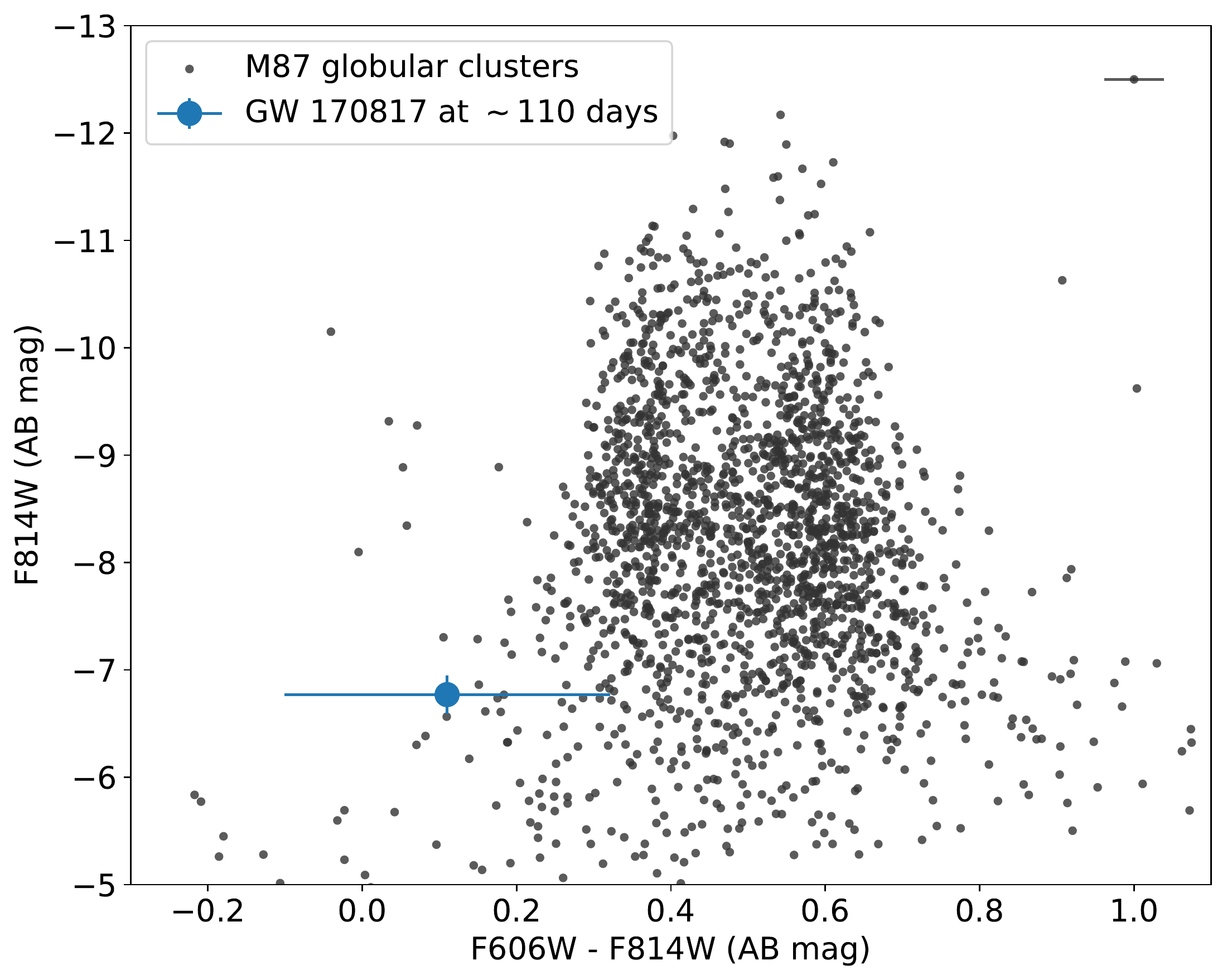}
\caption{{\bf \sffamily Colour-magnitude diagram of GW170817 at late times and globular clusters of M87.} We consider the possibility that our late-time detections are due to an underlying globular cluster. The optical colour of our detection of GW170817 (blue, large marker) appears bluer than the globular cluster population of early type galaxies (shown here for M87, small dark markers), although our large uncertainty on the colour means it cannot be significantly distinguished from a population of globular cluster colours. Uncertainties shown for GW170817 are 1$\sigma$. The globular cluster data\cite{peng09} have been trimmed to show only those with a 1$\sigma$ uncertainty in F606W -- F814W of $<$ 0.1\,mag -- individual uncertainty bars are not shown for clarity.}
\label{fig_gc}
\end{figure}

\begin{figure}
\centering
\includegraphics[width=\linewidth]{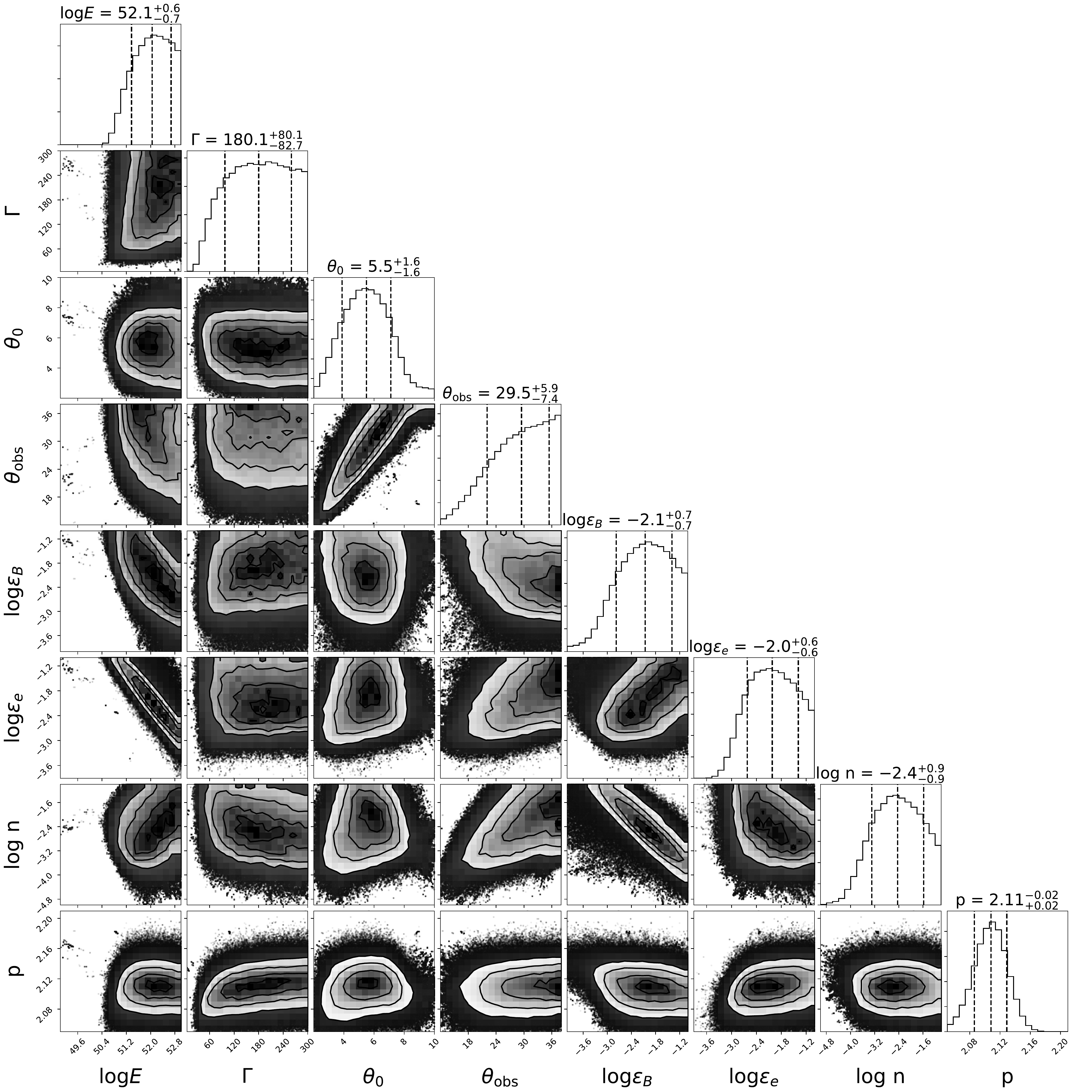}
\caption{{\bf \sffamily MCMC posterior distributions for a Gaussian structured jet model.} The parameter estimates for the structured jet model were found using an MCMC implementation fitting to the available multi-wavelength data. The parameter descriptions and units are given by the corresponding notation in Table \ref{modelparam}. Contours are used to denote the densest regions of the parameter space for each pair of parameters. For each parameter the 16, 50 and 84th percentiles of the distributions are indicated on the histograms; we take these as our best model values and uncertainties.}
\label{mcmc}
\end{figure}

Here we show that the available radio, optical and X-ray afterglow emission can be well modeled by a relativistic jet with Gaussian structured morphology\cite{lamb17}, chosen as a simple representation of a structured jet profile (see Supplementary Information). Model parameters producing good fits to the data were found using a Monte Carlo implementation, with the posteior parameters distrubtions shown in Figure~\ref{mcmc} and Table~\ref{modelparam}. The resulting light curve and spectra at the epochs with the best observational constraints on the afterglow are shown in Figure~\ref{modelfig} in comparison to our model. The continued rise of the radio and X-ray light curves from early to late times are well reproduced, and the late-time optical detections, near-IR limits and inferred radio-optical-X-ray spectral energy distribution we present here are also in good agreement with the model. The model suggests that the earlier optical and near-IR photometry was completely dominated by the kilonova light, with the contribution of the afterglow being $\gtrsim 29$\,mag (see Figure~\ref{lc_fig}). Unlike Ref.\,\cite{mooley17}, where the jet is  choked by the merger ejecta that powers a cocoon of material in the favored model, we find that the afterglow can be explained by an off-axis viewing angle of an initially highly relativistic jet with a bulk Lorentz factor $\Gamma\sim100$. At the observation time the jet is already decelerating and the components that contribute to the afterglow are less relativistic, $\Gamma\lesssim10$. If viewed close to on-axis, this jet would have an isotropic equivalent energy typical for other short GRBs (see Supplementary Information).  
Thus, our interpretation does not require the introduction of a new class of choked-jet events to explain GW170817, and instead makes this event consistent with arising from the same population of observed on-axis short GRBs.\cite{berger14}  

While our observations and modelling are consistent with an afterglow arising from a highly collimated jet, equally, we cannot rule out (or contradict) the existence of a cocoon.
The exact mechanism of how a jet develops it structure is unclear (as is jet formation itself). Ideal magneto-hydrodynamic
simulations show that neutron star mergers can self-consistently
produce jets along the binary rotation axis.\cite{ruiz16} In this setting, a jet forms and expands in a centrifugally
evacuated low-density funnel, therefore the jet may
not have significant interaction with any ejecta, and
may not result in a cocoon. The jet's structure in this case would
be due to the intrinsic formation and acceleration mechanism.
In nature, however, the jet will likely have to drill through earlier
ejected matter produced by neutrino-driven winds\cite{perego14} and/or shock-heated dynamical ejecta from the merger. This
interaction shapes the structure of the jet. In this scenario,
a cocoon will form that helps collimate the jet,\cite{bromberg11} and as it emerges from the ejecta it will gain structure due to the jet-cocoon interaction\cite{lazzati17}.

Existing observations do not allow us to determine the peak time of the afterglow light curve.  However, as long as the ambient density is at least $\sim 10^{-4}$ cm$^{-3}$ (a reasonable estimate, assuming the merger occurred within the interstellar or circumgalactic medium of the galaxy, and our modelling results, see SI), we expect that emission from the core of the jet will become visible within at most $t \sim 1$ year after the merger and possibly sooner (see Supplementary Information).  After the peak, the light curves across all frequency bands will plateau and then decay as a power law with an index between $-1.1$ and $-2.1$, depending on whether the jet spreads sideways and when it ceases to be relativistic.  This prediction is distinct from the mildly relativistic cocoon model, which predicts a longer rise and ultimately a shallower decay, as $t^{-0.8}$.  We thus anticipate that it will be possible to determine the correct model in the near future.

Afterglows from jets with a structure that extends beyond a narrow `top-hat' shaped core can peak earlier and brighter at modest inclinations, or have an early excess that results in a more gradually rising afterglow, than a purely `top-hat' shaped jet.
Therefore they imply an increase in the rate of orphan afterglows for deep optical surveys, $\gtrsim 23$ mag.\cite{lamb18}   
Assuming similar jet parameters to those estimated for GW170817, we expect short GRBs with energies typical of cosmological bursts (isotropic equivalent energies $\gtrsim 3\times10^{49}$ erg) to be associated with $\sim$ 5---15\% of binary neutron star gravitational-wave detections (cf. Ref \cite{lazzati17}). This accounts for gravitational-wave selection effects, which moderately favour nearly on-axis sources.  The effective opening angle at which a structured jet such as the one proposed here could be observed at cosmological distances, $\sim 10$ degrees, yields a beaming factor of $\sim 100$, broadly consistent with that inferred by observations.\cite{fong12} This yields a true estimate of short GRB rates of order $1000$ Gpc$^{-3}$ yr$^{-1}$, consistent with Galactic double neutron star observations\cite{abbott10} and the rate inferred from GW170817.\cite{lvc17a}

In this model, an observer viewing down the axis of the merger would have observed a short GRB that is comparable in energy to those typically seen by $\gamma$-ray satellites.  
On the other hand, an observer at a higher inclination would be in a part of the jet with a lower energy that could result in a sub-luminous GRB such as the one seen associated with GW170817.\cite{granot17grb,lamb17b}
The range of jet energies directed toward the observer in structured jets could explain the puzzling diversity in observed short GRB luminosities.\cite{berger14}

\begin{addendum}
 \item 
 Based on observations made with the NASA/ESA Hubble Space Telescope, obtained from the data archive at the Space Telescope Science Institute. STScI is operated by the Association of Universities for Research in Astronomy, Inc. under NASA contract NAS 5-26555. These observations are associated with programs GO 14771 (Tanvir), GO 14270 (Levan). We thank the staff at STScI for their excellent support of these observations. AJL acknowledges that this project has received funding from the European Research Council (ERC) under the European Union's Horizon 2020 research and innovation programme (grant agreement no 725246).
JDL, AJL, DS, KW acknowledge support from STFC via grant ST/P000495/1. NRT, PTO, JPO,  
GPL, IM, SK acknowledge support from STFC. GPL acknowledges partial support from RAS and IAU grants.
JH was supported by a VILLUM FONDEN Investigator grant (project number 16599). The Cosmic Dawn Center is funded by the DNRF. 
AdUP, CT and ZC acknowledge support from the Spanish project AYA 2014-58381-P.  ZC also acknowledges support from the Juan de la Cierva Incorporaci\'on fellowship IJCI-2014-21669.
MI acknowledges the support from the NRFK grant, No. 2017R1A3A3001362.
SR has been supported by the Swedish Research Council (VR) under grant number 2016- 03657\_3, by the Swedish National Space Board under grant number Dnr. 107/16 and by the research environment grant ``Gravitational Radiation and Electromagnetic Astrophysical Transients (GREAT)" funded by the Swedish Research council (VR) under Dnr 2016-06012. 
PAE acknowledges UKSA support. DJW is supported by the the Danish Agency for Science, Technology and Innovation under grant number DFF – 7014-00017.
GPL thanks Adam Higgins and Liam Raynard for useful conversations regarding MCMC, and GPL and SK thank Ehud Nakar for helpful comments.
IM thanks Jonathan Granot for useful discussions.
 \item [Author Contributions] JDL performed the data reduction and analysis and led writing of the manuscript. GPL performed the numerical calculations and wrote text relating to the model. IM contributed to theoretical interpretation of the data and provided analytical estimates. AJL and NRT are PIs of the {\em HST} proposals used to obtain the new data presented and assisted with data analysis and text. SK assisted with the development of the model. BG and JH contributed to the interpretation of the data and performed the phenomenological fits. ASF and TK performed the image subtraction test. All authors provided comments and analysis to assist in the writing of the observing proposals and manuscript.
 \item[Competing Interests] The authors declare that they have no competing interests.
  \item[Corresponding Author] Correspondence should be addressed to JDL (email: J.D.Lyman@warwick.ac.uk).
\end{addendum}

\begin{table}
\centering
 \begin{tabular}{cccccc} 
 \hline
 Obs. date & Rest-frame epoch & Instrument & Filter & Exp. time & AB Mag \\
   (MJD)          & (days)          &            &        & (s)           &        \\
 \hline
 58093.007 & 109.41 & WFC3/UVIS & F606W & 2264 & 26.40 $\pm$ 0.11 \\
 58093.074 & 109.47 & WFC3/IR & F160W & 2397 & $>$25.0 \\
 58093.139 & 109.54 & WFC3/UVIS & F814W & 2400 & 26.29 $\pm$ 0.18 \\
 58093.206 & 109.61 & WFC3/IR & F140W & 4794 & $>$25.3  \\
 \hline
 \end{tabular}
  \caption{{\bf \sffamily Log of late time {\em HST} observations of GW170817.} Epochs for observations are given as time since the gravitational wave signal (correct for the source's redshift of z = 0.007983). Magnitudes have been corrected for Galactic extinction using E(B--V) = 0.105\,mag. Uncertainties are 1$\sigma$ and limits in the IR channel are given as 3$\sigma$.}
  \label{obslog}
\end{table}

\begin{methods}
\subsection{Observations and reductions.}
Observations were taken with Wide Field Camera 3 (WFC3) onboard {\em Hubble Space Telescope} using both UVIS (F606W and F814W) and IR (F140W and F160W) channels. Each set of exposures was  divided into 4 dithered pointings to allow for the use of image drizzling\cite{fruchter02} in order to improve the spatial resolution. All processed frames were obtained from the {\em HST} data archive and drizzled to pixel scales of 0.025$''$ and 0.07$''$ for the UVIS and IR channels, respectively.

\subsection{Astrometry and photometry.}
We determined the location of GW170817 in the images via relative astrometry using our {\em HST} observations taken at earlier epochs.\cite{tanvir17,levan17} Using 20--30 point sources in common to each pair of images, the geometric transformations achieve an rms~$\sim$~0.1--0.2~pixels in each filter. We do not have an early epoch of F140W image, and so this was tied to our earlier F110W image. Since the source lies at a small projected offset from its bright host galaxy, photometry requires the removal of the galaxy light. In principle this can be done simply by estimating the background light in a small aperture around the source position. However, in order to subtract the gradient of the galaxy light, we modelled the galaxy as a series of elliptical isophotes with the IRAF task {\tt ellipse}. Prior to creating this model, point sources in the images were masked after automated detection with {\tt SExtractor}.\cite{bertin96} The removal of such sources is required in order to afford an accurate determination of the galaxy's underlying surface brightness profile when constructing the model.
This isophotal model was subtracted from our images to yield a frame in which the galaxy background is removed. Although residuals were present in the inner regions where the morphology of NGC~4993 is complex\cite{levan17}, at the location of GW170817 the background was smoothly subtracted. Aperture photometry was then performed in 0.08 and 0.15$''$ radii apertures (for UVIS and IR channels, respectively) using the local background to estimate the uncertainty and corrected according to the published WFC3 encircled energy curves. Finally we corrected for Galactic extinction in the direction of NGC~4993 of $E(B-V) = 0.105$\,mag\cite{schlafly11} using an $R = 3.1$ extinction law.\cite{fitzpatrick99} As a check of the method, we also made use of publicly available pre-merger {\em HST} imaging in order to perform image subtraction\cite{alard00} to remove the galaxy background light. Pre-merger imaging is only available in F606W and is shallow (exposure time 696~seconds) compared to our imaging. After aligning and subtracting this template galaxy image we repeated our photometry method and found $m_\text{F606W} = 26.44 \pm 0.14$\,mag. This image subtraction magnitude is in excellent agreement with our value determined by subtracting an elliptical isophote model as above ($m_\text{F606W} = 26.40 \pm 0.11$\,mag), with a larger uncertainty due to the use of the comparatively noisy template image.  We therefore report magnitudes in both F606W and F814W based on the removal of the model galaxy.

\end{methods}

\begin{addendum}
 \item [Data Availability] The newly-presented \textit{HST} data are stored in the Mikulski Archive for Space Telescopes (https://archive.stsci.edu/hst/) and available from the corresponding author upon reasonable request.
 \item [Code Availability] The algorithm for the structured jet model used here is fully described in Ref.\cite{lamb17} and the MCMC implementation was done via the publicly available emcee\cite{emcee} package (http://dfm.io/emcee/current/). The specific codes are available upon request to the corresponding author.

\end{addendum}

\begin{supplinfo}

\subsection{Optical detections as being due to an underlying, unrelated source.}
Given the depth of our observations, and the relative proximity of NGC~4993, the absolute magnitude we are sensitive to at the location of GW170817 ($\gtrsim -6$\,mag) probes well down the luminosity function of typical globular cluster (GC) or young star cluster populations (e.g. the average Galactic GC luminosity is $\sim -8$\,mag\cite{harris96}). As such we consider the possibility that our detections are due to an unrelated source. Firstly, we assessed the presence of any offset between our new detections and earlier {\em HST} epoch detections of the kilonova light. We were able to determine a relative astrometric transformation with a 0.1--0.2 pixels r.m.s.\ ($\sim 2.5-5$\,mas, or $\sim 0.5-1$\,pc at the distance of NGC~4993). However, owing to the relative faintness of our new detections, our main source of uncertainty was the centroiding of the source itself. Although this can be estimated as $\frac{\text{FWHM}}{2.35 \times \text{SNR}}$, practically we were also subject to uncertainties based on our choice of method to centre the faint source. The centre uncertainties were $\sim 2-3$\,pc (the uncertainty in the centre of the source in our earlier epochs was negligible owing to the much higher significance of those detections).
When considering our sources of uncertainty, we find the new sources have no significant offset from our earlier epoch detections. 
Secondly, we consider the colour of the new detections. The source is relatively blue, at F606W\,$-$\,F814W\,$= 0.11 \pm 0.21$\,mag, when compared to GCs, which typically have $\sim0.5$\,mag in the same colour.\cite{madrid09,carlson17} However, the large uncertainty on the colour of our detection precludes strong statements when comparing to colours of GCs. In Figure~\ref{fig_gc} we show the GC population of the nearby elliptical galaxy M87 ($D_L \simeq 16.5$\,Mpc) in comparison to our detection.\cite{peng09}
We also find a lack of other similarly blue sources within NGC~4993 -- this is consistent with the S0 classification and studies showing a lack of recent star-formation in the galaxy.\cite{blanchard17,im17,levan17} 
Additionally, we note that no source was found in the pre-existing {\em HST} imaging of the merger site in F606W, although these data were shallower in depth than our observations.\cite{levan17, blanchard17,smartt17} 
Our F606W subtraction against the pre-imaging of the galaxy (see Methods) still reveals the source, suggesting that it was not present in the pre-imaging observations, even at a marginal level, and supports its interpretation as a transient source
Although binary neutron-stars may be dynamically formed in dense stellar clusters,\cite{davies95} the natal kicks imparted to the neutron stars at birth are significantly larger than the escape velocity of such clusters (few tens of km~s$^{-1}$). As such, there is no reason to expect the presence of a stellar cluster underlying the merger site (above that expected for a random location at similar offset in the host), even in the dynamically-formed scenario.
Given the above arguments and the consistency of our detections with power-law models of the afterglow emission from radio and X-ray data, we consider that our detections are indeed due to the transient for the purposes of this study. However, continued long-term optical monitoring of the source alongside radio and X-ray observations is required to conclusively rule on whether the emission is due to an unrelated, underlying source. In such case we would not expect any significant change in the observed flux, unlike the evolving optical light curve expected for an afterglow model.

\section*{Off-axis afterglow constraints.}

Here, we summarize the basic constraints that can be placed on the system parameters from the afterglow model.  Our detailed model of a structured jet is presented in the following section. 

Observationally, the emission from GW170817 at $\sim108-110$~days after merger is broadly described by a spectral power law extending from radio through optical to X-ray frequencies, with $F_\nu \propto \nu^{-0.55}$ to $\nu^{-0.6}$.
This spectral slope is consistent with that found from radio and X-ray observations at earlier times, although the uncertainties at $\sim 15$ days after merger are much larger.  
A spectral power-law with a slope of $-0.55$ to $-0.6$ below the cooling break is consistent with a GRB afterglow with an electron energy distribution power-law slope $p \sim 2.1$ to $2.2$.\cite{saripirannarayan1998, granotsari02}

This robust power-law spectrum implies that all observational frequencies are within or close to the range between the synchrotron frequency of minimal energy electrons (which must be below the radio frequency, $\nu_m < 3$ GHz) and the cooling frequency (which must be above or not much below the X-ray frequency, $\nu_c \gtrsim 2.4 \times 10^{17}$ Hz).  However, $\nu_m$ and $\nu_c$ depend on a number of parameters, including fractions of the internal energy in the electrons $\epsilon_e$ and in the magnetic field $\epsilon_B$, and this requirement does not significantly constrain these beyond $\epsilon_B \lesssim 0.1$, $\epsilon_e \lesssim 0.01$.  

The lack of a prompt X-ray signature,\cite{evans17} the dimness of the GRB itself,\cite{lvc17grb} and the brightness and continuing rise of the radio and X-ray afterglow,\cite{hallinan17,kim17,mooley17,troja17,margutti17,ruan17} indicate any relativistic jet, launched as part of the merger, is being viewed off-axis. {Depending on the jet's morphology, prompt gamma-ray signals can be observed even for significantly off-axis events.\cite{kathirgamaraju18}} However, the observational constraints have also been used to argue against the afterglow being due to a jet, instead possibly pointing to a more spherical cocoon of mildly-relativistic material as the source.\cite{kasliwal17,hallinan17,gottlieb17, mooley17} This cocoon of material would be formed due to interaction between the jet and the ejecta from the merger. This merger interaction is favoured to have choked the jet in current cocoon interpretations for GW170817, resulting in no escape of a highly relativistic jet.
For binary neutron star mergers, however, the amount of ejecta is small (a few hundredths of a solar mass was inferred for GW170817\cite{kasen17}), 
with a significant fraction in the binary plane,\cite{metzger10,kasen17,tanaka17,shibata17} away from the polar direction of the jet. Therefore, it is not clear whether the jet can be choked or power a significant cocoon.\cite{lazzati17}
 
However, when considering the off-axis jet model, the relatively flat rise of the afterglow light curve argues against a single `top-hat' jet with sharp edges.  A top-hat jet would have a $t^3$ rise before peak luminosity, much steeper than the observed rise which scales at most linearly with time between $\sim 10$ and $\sim 100$ days after the merger.  This argument has already been made on other grounds, such as the GRB itself: the relatively short delay time between the gravitational-wave signal and the GRB is inconsistent with moderately off-axis viewing unless the gamma-rays are emitted at very small radii, and the relationship between the observed gamma-ray flux and typical photon energies point to a relatively low-energy portion of the outflow being directed at the observer\cite{granot17grb} (cf. Ref.\,\cite{kisaka17}).

Whether the jet angular energy distribution is a top-hat or a Gaussian, as we consider in the following section, the observed flux will peak at a time when the observer located at angle $\theta_\mathrm{obs}$ off-axis will be able to see the jet from the energetic on-axis core, i.e., when the jet core's Lorentz factor drops to $\Gamma \sim 1/\theta_\mathrm{obs}$.  If the jet does not expand sideways, the light curve peaks at
\begin{equation}
t_\mathrm{peak} \approx 160 \left(\frac{n}{10^{-4}\,\mathrm{cm}^{-3}}\right)^{-1/3} \left(\frac{E_K}{10^{50}\,\mathrm{erg}}\right)^{1/3} \left(\frac{\theta_\mathrm{obs}}{20\,\mathrm{deg}}\right)^2
\left(\frac{\theta_\mathrm{obs}}{\theta_0}\right)^{2/3}~\mathrm{days}\,,
\end{equation}
where $E_K$ is the total explosion energy of the system, $n$ is the ambient density, and $\theta_0$ is the initial jet core opening angle.  This can be expressed in terms of the isotropic-equivalent energy of the jet core, $E_0=2 E_K / \theta_0^2$, as
\begin{equation}
t_\mathrm{peak} \approx 420 \left(\frac{n}{10^{-4}\,\mathrm{cm}^{-3}}\right)^{-1/3} \left(\frac{E_0}{3\times 10^{52}\,\mathrm{erg}}\right)^{1/3} \left(\frac{\theta_\mathrm{obs}}{20\,\mathrm{deg}}\right)^{8/3}
~\mathrm{days}\,.
\label{Eq.3}
\end{equation}

Here, we used fiducial values corresponding to the maximum isotropic-equivalent energy of observed on-axis GRBs and close to minimal halo ambient density\cite{fong15}.  The maximum allowed inclination angle based on gravitational-wave data, the most recent Hubble constant measurements from the Dark Energy Survey, and the known redshift of the host galaxy NGC~4993 is $\theta_\mathrm{obs}\le 28^\circ$ at 90\% confidence.\cite{lvc17a,mandel17}.  Using this maximum observing angle extends the afterglow peak to $\sim 1000$ days.\cite{mandel17}  Numerical simulations indicate that the light curves could peak somewhat later than in the analytical treatment.\cite{granot17a}  On the other hand, 
it has been suggested\cite{granotsari02} that these expressions should use $\Delta \theta \equiv \theta_\mathrm{obs} - \theta_0$, the angle by which the observer is outside the jet core, in lieu of $\theta_\mathrm{obs}$, reducing the peak time.  Therefore, we conclude that the light curve should reach the peak within $\sim 1$ year after the merger.  Subsequently, the light curve from the post-jet-break off-axis jet will decay as $t^{-p} \sim t^{-2.1}$ if the jet expands sideways\cite{granot17a} or more slowly, as $\sim t^{3(1-p)/4-3/4} \sim t^{-1.6}$ if it does not.\cite{panaitescu99}  The light curve decay will flatten to $\sim t^{-1.1}$ when the jet becomes non-relativistic \cite{frailwaxmankulkarni2000,sironigiannios2013} and the outflow transitions to a Sedov-Taylor solution; it may therefore never reach the steeper $t^{-1.6}$ -- $t^{-2.1}$ decline if this transition happens soon after the peak.
This is in contrast to the cocoon model, where there would be a longer continued rise and shallower decay afterwards, as $t^{3(1-p)/4} \sim t^{-0.8}$ in the relativistic regime, consistent with near-spherical ejecta.\cite{granotsari02} Continued monitoring of the source to very late times will distinguish between the two scenarios.

\subsection{The structured jet afterglow model.}
Here we consider a Gaussian structured jet as a model for the afterglow of GW170817; further details of the model are given in Ref \cite{lamb17}. The energy per solid angle and $\Gamma_0-1$, where $\Gamma_0 \equiv \Gamma(t=0)$ is the bulk Lorentz factor in the coasting phase before deceleration, vary with angle from the central axis as $\propto e^{-{\theta^2/2\theta_0^2}}$. Here $\theta$ is the angle from the central axis and $\theta_0$ is the angular scale that defines the jet core.
Note that this corresponds to a low Lorentz factor for the portion of the jet directed toward the observer, which would therefore likely be opaque to prompt gamma rays with a synchrotron origin from the dissipation radius\cite{kisaka17}; this may indicate a shallower Lorentz factor distribution across the jet or that the gamma rays are emitted at the photosphere.
The minimum Lorentz factor for producing the gamma ray energy, considering emission from material travelling towards an observer, is $\Gamma\sim 8$ [ref. \cite{lamb2016}], resulting in the gamma rays being beamed into an emission cone with an angle $\sim 7^\circ$. In this case the dissipation radius would not be sufficiently below the photosphere to supress the emission of the gamma-rays.
The observed prompt emission could therefore have been viewed off-axis from the slower and less energetic wider component of a structured jet or, alternatively, may have been produced by another mechanism such as the breakout of a shock produced by the jet's passage through the dynamical ejecta.

The numerical model splits the jet into a number of segments with a solid angle $\Omega$.
The flux contribution from each segment is calculated for an observer at an angle $i$ from that segment's central axis.
For an on-axis observer the segment flux is 
\begin{equation}
F_\nu(t)=(L_\nu/4\pi D_L^2)(\Omega/\Omega_e),
\end{equation}
where $L_\nu$ is the luminosity, $D_L$ the luminosity distance, and $\Omega_e= {\rm max}[\Omega,~\Omega_\Gamma]$. Here $\Omega_\Gamma=2\pi(1-\cos[1/\Gamma(\theta,t)])$ with $t$ the time after deceleration.
The flux from each of these small segments can be treated as a point source.  For an off-axis observer at an inclination $\theta_{\rm obs}$ from the jet central axis, or $i$ from an individual jet segment, the flux becomes $F_\nu(t,i)=a^3 F_{\nu/a}(at,\ i=0)$, where $a=\delta(i)/\delta(i=0)<1$, $\delta$ is the relativistic Doppler factor $[\Gamma(1-\beta\cos i)]^{-1}$, and $\beta \equiv \sqrt{1-\Gamma^{-2}}$.

There are a number of free parameters in the model, and the data do not constrain a unique solution.  To assess the best fitting model, and degeneracies between parameters, we use an MCMC\cite{emcee} implementation of the model. Our prior ranges are uniform (uniform in log where the log of the value is shown in Figure \ref{mcmc}) over reasonable parameter values, guided by external considerations based on observations of GRB samples\cite{fong12, fong15} as well as constraints on the inclination from the GW and electromagnetic signals. The ensemble MCMC run consisted of 300 walkers taking 10000 steps. In Figure \ref{mcmc} we show the corner plot for our MCMC results. Our best fitting parameters are shown in Table \ref{modelparam}. We note the bulk Lorentz factor of the jet is completely unconstrained in this model, and we can only state that we require $\Gamma_0 > 60$; 
models with low values of $\Gamma_0$ fail to reproduce the observations at $\sim10-20$ days due to the longer deceleration time-scale for low-$\Gamma$ outflows. The GW signal constrained the merger to be inclined $< 55$~degrees\cite{lvc17a}, with a combined GW and electromagnetic constraint at 90\% confidence of $< 28$~degrees\cite{mandel17}, although we note this value is dependent on the choice of $H_0$. Our inclination of $29.5\substack{+5.9\\-7.4}$~degrees is in good agreement with these constraints.
The energy is consistent with that seen for cosmological GRBs, where a third of the short GRB population has an isotropic jet kinetic energy larger than $\sim 10^{52}$ erg,\cite{fong15} the micro-physical parameters $\varepsilon_B$ and $\varepsilon_e$ are comfortably in the expected range,\cite{medvedev2006} and the ambient density is reasonable for the location of GW170817 within the host galaxy.\cite{oppenheimer16}

We show in Figure 3 of the main text our structured jet model constructed from the median of our posterior resultant light curves and include a shaded region that indicates the range of diversity in the light curves for the parameter distributions.
The light curve and spectra were extracted at $\sim16$ and $\sim110$ days to be roughly contemporaneous with the available afterglow data.
The model spectra at $\sim 16$ days is a single power law with an index $\sim-0.56$, however a break due to the cooling frequency can be seen at $\sim3\times10^{17}$ Hz. At 16 days, radio, optical and X-ray  emission is between the synchrotron characteristic frequency, which must be $<3\times10^9$ Hz, and the cooling frequency.
At $\sim 110$ days, a break at $\sim3\times 10^{16}$ Hz indicates the passage of the cooling frequency through the X-ray band. 
The index below the break is $-0.56$ and above the break $\sim-1.06$.

The first break in the model light-curve is the jet deceleration time for an off-axis observer, $t \propto E_0^{1/3} n^{-1/3} \Gamma_0^{-8/3} a^{-1}$, where before the break the flux is $\propto t^3$ and the jet is assumed not to expand sideways.
A higher jet core Lorentz factor moves this break to earlier times. 
Each jet component will have a beaming angle equal to a given angle $i$ at post-merger time \\ $t \propto E_0^{1/3} n^{-1/3} i^{8/3} $ for an on-axis observer, $\theta_{\rm obs}=0$, or at a factor $a^{-1} = (1-\beta\cos i)/(1-\beta)\sim 2$ of the on-axis time for an off-axis observer at an inclination $i=\Gamma^{-1}$.
Here $E_0=2 E_K \theta_0^{-2}$ is the isotropic equivalent jet kinetic energy of the component, and $n$ is the ambient number density.
For wider jet components with a low Lorentz factor, emission can be beamed towards an observer before the jet starts to decelerate.
After this time, flux from the more energetic and faster components is beamed towards the observer. 

We do not yet know when the emission will reach its peak; a `top-hat' jet with the core parameters of the Gaussian jet model and without lateral expansion or limb brightening, would yield a peak at $\sim 260$ days for an observer at the model inclination of $\theta_\mathrm{obs}=29.5^\circ$.
Lateral expansion will reduce the peak time, while significant limb brightening will add to the flux after the peak for off-axis observers.
The late time flux decline approaches the `on-axis' post jet-break flux, for our parameters, without sideways expansion, as $\propto t^{-1.6}$.
At very late times when the jet becomes non-relativistic the decline will transition from the steep Blandford-McKee solution to a Sedov-Taylor solution, $t^{-1.1}$. 
At this point the receding counter-jet will become observable, resulting in a bump in the afterglow decline.
Late-time, $\sim 2$ years post-merger, radio observations may be able to reveal this feature.

\end{supplinfo}

\newpage

\begin{table}
\centering
\begin{tabular}{c c c}
 Model Parameter & Description & Value \\
\hline
$E_0$        & Isotropic equivalent kinetic energy & $1.26\substack{+3.75\\-1.01} \times10^{52}$\, erg \\
$\Gamma_0$           & Initial bulk Lorentz factor         & $180.1\substack{+80.1\\-82.7}$ \\
$n$                & Ambient density                     & $0.40\substack{+2.76\\-0.35}\times10^{-2}$\,cm$^{-3}$ \\
$\varepsilon_e$    & Electron energy fraction            & $1.0\substack{+3.0\\-0.8} \times 10^{-2}$ \\
$\varepsilon_B$    & Magnetic field energy fraction      & $0.8\substack{+3.2\\-0.7} \times10^{-2}$ \\
$p$                & Electron energy distribution        & $2.11\substack{+0.02\\-0.02}$ \\
$\theta_0$         & Opening angle of jet core           & $5.5\substack{+1.6\\-1.6}$ deg \\
$\theta_{\rm obs}$ & Observing angle                     & $29.5\substack{+5.9\\-7.4}$ deg \\
$D_L$              & Luminosity distance                 & $41$\,Mpc \\
\end{tabular}
\\
\caption{{\bf \sffamily Parameters for Gaussian structured jet.} 
 Best model parameters and their uncertainty correspond to the 16, 50 and 84th percentile on the MCMC posterior distributions of the parameters. The values for $E_0$ and $\Gamma_0$ are the point values for the centre of the jet core. Some of the parameters are significantly correlated, and some are largely unconstrained by current data; e.g., the bulk Lorentz factor is only constrained to $\Gamma_0 \geq 60$ and thus the mean value and uncertainty are driven almost entirely by our prior range. The value of $D_L$ was fixed based on other electromagnetic observations.}
\label{modelparam}
\end{table}

\newpage 
\newcommand{\araa}{Annu. Rev. Astron. Astr.}   \newcommand{\aap}{Astron. Astrophys.}
\newcommand{\aj}{Astron. J.}         \newcommand{\apj}{Astrophys. J.}
\newcommand{\apjl}{Astrophys. J. Lett.}      \newcommand{\apjs}{Astrophys. J. Suppl. S.}
\newcommand{\mnras}{Mon. Not. R. Astron. Soc.}   \newcommand{\nat}{Nature}
\newcommand{\pasp}{Publ. Astron. Soc. Pac.}  \newcommand{\pasj}{Publ. Astron. Soc. Jpn.}
\newcommand{\aaps}{Astron. Astrophys. Supp.}
\newcommand{\prd}{Phys. Rev. D}

\bibliography{references}

\begin{thebibliography}{10}
\expandafter\ifx\csname url\endcsname\relax
  \def\url#1{\texttt{#1}}\fi
\expandafter\ifx\csname urlprefix\endcsname\relax\def\urlprefix{URL }\fi
\providecommand{\bibinfo}[2]{#2}
\providecommand{\eprint}[2][]{\url{#2}}

\bibitem{lvc17a}
\bibinfo{author}{{Abbott}, B.~P.} \emph{et~al.}
\newblock \bibinfo{title}{{GW170817: Observation of Gravitational Waves from a
  Binary Neutron Star Inspiral}}.
\newblock \emph{\bibinfo{journal}{Physical Review Letters}}
  \textbf{\bibinfo{volume}{119}}, \bibinfo{pages}{161101}
  (\bibinfo{year}{2017}).

\bibitem{lvc17b}
\bibinfo{author}{{Abbott}, B.~P.} \emph{et~al.}
\newblock \bibinfo{title}{{Multi-messenger Observations of a Binary Neutron
  Star Merger}}.
\newblock \emph{\bibinfo{journal}{\apjl}} \textbf{\bibinfo{volume}{848}},
  \bibinfo{pages}{L12} (\bibinfo{year}{2017}).

\bibitem{lvc17grb}
\bibinfo{author}{{Abbott}, B.~P.} \emph{et~al.}
\newblock \bibinfo{title}{{Gravitational Waves and Gamma-Rays from a Binary
  Neutron Star Merger: GW170817 and GRB 170817A}}.
\newblock \emph{\bibinfo{journal}{\apjl}} \textbf{\bibinfo{volume}{848}},
  \bibinfo{pages}{L13} (\bibinfo{year}{2017}).

\bibitem{arcavi17}
\bibinfo{author}{{Arcavi}, I.} \emph{et~al.}
\newblock \bibinfo{title}{{Optical emission from a kilonova following a
  gravitational-wave-detected neutron-star merger}}.
\newblock \emph{\bibinfo{journal}{\nat}} \textbf{\bibinfo{volume}{551}},
  \bibinfo{pages}{64--66} (\bibinfo{year}{2017}).

\bibitem{coulter17}
\bibinfo{author}{{Coulter}, D.~A.} \emph{et~al.}
\newblock \bibinfo{title}{{Swope Supernova Survey 2017a (SSS17a), the optical
  counterpart to a gravitational wave source}}.
\newblock \emph{\bibinfo{journal}{Science}} \textbf{\bibinfo{volume}{358}},
  \bibinfo{pages}{1556--1558} (\bibinfo{year}{2017}).

\bibitem{evans17}
\bibinfo{author}{{Evans}, P.~A.} \emph{et~al.}
\newblock \bibinfo{title}{{Swift and NuSTAR observations of GW170817: Detection
  of a blue kilonova}}.
\newblock \emph{\bibinfo{journal}{Science}} \textbf{\bibinfo{volume}{358}},
  \bibinfo{pages}{1565--1570} (\bibinfo{year}{2017}).

\bibitem{pian17}
\bibinfo{author}{{Pian}, E.} \emph{et~al.}
\newblock \bibinfo{title}{{Spectroscopic identification of r-process
  nucleosynthesis in a double neutron-star merger}}.
\newblock \emph{\bibinfo{journal}{\nat}} \textbf{\bibinfo{volume}{551}},
  \bibinfo{pages}{67--70} (\bibinfo{year}{2017}).

\bibitem{smartt17}
\bibinfo{author}{{Smartt}, S.~J.} \emph{et~al.}
\newblock \bibinfo{title}{{A kilonova as the electromagnetic counterpart to a
  gravitational-wave source}}.
\newblock \emph{\bibinfo{journal}{\nat}} \textbf{\bibinfo{volume}{551}},
  \bibinfo{pages}{75--79} (\bibinfo{year}{2017}).

\bibitem{soares17}
\bibinfo{author}{{Soares-Santos}, M.} \emph{et~al.}
\newblock \bibinfo{title}{{The Electromagnetic Counterpart of the Binary
  Neutron Star Merger LIGO/Virgo GW170817. I. Discovery of the Optical
  Counterpart Using the Dark Energy Camera}}.
\newblock \emph{\bibinfo{journal}{\apjl}} \textbf{\bibinfo{volume}{848}},
  \bibinfo{pages}{L16} (\bibinfo{year}{2017}).

\bibitem{tanvir17}
\bibinfo{author}{{Tanvir}, N.~R.} \emph{et~al.}
\newblock \bibinfo{title}{{The Emergence of a Lanthanide-rich Kilonova
  Following the Merger of Two Neutron Stars}}.
\newblock \emph{\bibinfo{journal}{\apjl}} \textbf{\bibinfo{volume}{848}},
  \bibinfo{pages}{L27} (\bibinfo{year}{2017}).

\bibitem{valenti17}
\bibinfo{author}{{Valenti}, S.} \emph{et~al.}
\newblock \bibinfo{title}{{The Discovery of the Electromagnetic Counterpart of
  GW170817: Kilonova AT 2017gfo/DLT17ck}}.
\newblock \emph{\bibinfo{journal}{\apjl}} \textbf{\bibinfo{volume}{848}},
  \bibinfo{pages}{L24} (\bibinfo{year}{2017}).

\bibitem{haggard17}
\bibinfo{author}{{Haggard}, D.} \emph{et~al.}
\newblock \bibinfo{title}{{A Deep Chandra X-Ray Study of Neutron Star
  Coalescence GW170817}}.
\newblock \emph{\bibinfo{journal}{\apjl}} \textbf{\bibinfo{volume}{848}},
  \bibinfo{pages}{L25} (\bibinfo{year}{2017}).

\bibitem{hallinan17}
\bibinfo{author}{{Hallinan}, G.} \emph{et~al.}
\newblock \bibinfo{title}{{A radio counterpart to a neutron star merger}}.
\newblock \emph{\bibinfo{journal}{Science}} \textbf{\bibinfo{volume}{358}},
  \bibinfo{pages}{1579--1583} (\bibinfo{year}{2017}).

\bibitem{mooley17}
\bibinfo{author}{{Mooley}, K.~P.} \emph{et~al.}
\newblock \bibinfo{title}{{A mildly relativistic wide-angle outflow in the
  neutron-star merger event GW170817}}.
\newblock \emph{\bibinfo{journal}{\nat}} \textbf{\bibinfo{volume}{554}},
  \bibinfo{pages}{207--210} (\bibinfo{year}{2018}).

\bibitem{ruan17}
\bibinfo{author}{{Ruan}, J.~J.}, \bibinfo{author}{{Nynka}, M.},
  \bibinfo{author}{{Haggard}, D.}, \bibinfo{author}{{Kalogera}, V.} \&
  \bibinfo{author}{{Evans}, P.}
\newblock \bibinfo{title}{{Brightening X-Ray Emission from GW170817/GRB
  170817A: Further Evidence for an Outflow}}.
\newblock \emph{\bibinfo{journal}{\apjl}} \textbf{\bibinfo{volume}{853}},
  \bibinfo{pages}{L4} (\bibinfo{year}{2018}).

\bibitem{kasen17}
\bibinfo{author}{{Kasen}, D.}, \bibinfo{author}{{Metzger}, B.},
  \bibinfo{author}{{Barnes}, J.}, \bibinfo{author}{{Quataert}, E.} \&
  \bibinfo{author}{{Ramirez-Ruiz}, E.}
\newblock \bibinfo{title}{{Origin of the heavy elements in binary neutron-star
  mergers from a gravitational-wave event}}.
\newblock \emph{\bibinfo{journal}{\nat}} \textbf{\bibinfo{volume}{551}},
  \bibinfo{pages}{80--84} (\bibinfo{year}{2017}).

\bibitem{waxman17}
\bibinfo{author}{{Waxman}, E.}, \bibinfo{author}{{Ofek}, E.},
  \bibinfo{author}{{Kushnir}, D.} \& \bibinfo{author}{{Gal-Yam}, A.}
\newblock \bibinfo{title}{{Constraints on the ejecta of the GW170817
  neutron-star merger from its electromagnetic emission}}.
\newblock \emph{\bibinfo{journal}{ArXiv e-prints}}  (\bibinfo{year}{2017}).

\bibitem{rosswog17}
\bibinfo{author}{{Rosswog}, S.} \emph{et~al.}
\newblock \bibinfo{title}{{Detectability of compact binary merger macronovae}}.
\newblock \emph{\bibinfo{journal}{Classical and Quantum Gravity}}
  \textbf{\bibinfo{volume}{34}}, \bibinfo{pages}{104001}
  (\bibinfo{year}{2017}).

\bibitem{drout17}
\bibinfo{author}{{Drout}, M.~R.} \emph{et~al.}
\newblock \bibinfo{title}{{Light curves of the neutron star merger
  GW170817/SSS17a: Implications for r-process nucleosynthesis}}.
\newblock \emph{\bibinfo{journal}{Science}} \textbf{\bibinfo{volume}{358}},
  \bibinfo{pages}{1570--1574} (\bibinfo{year}{2017}).

\bibitem{kumargranot03}
\bibinfo{author}{{Kumar}, P.} \& \bibinfo{author}{{Granot}, J.}
\newblock \bibinfo{title}{{The Evolution of a Structured Relativistic Jet and
  Gamma-Ray Burst Afterglow Light Curves}}.
\newblock \emph{\bibinfo{journal}{\apj}} \textbf{\bibinfo{volume}{591}},
  \bibinfo{pages}{1075--1085} (\bibinfo{year}{2003}).

\bibitem{zhang04}
\bibinfo{author}{{Zhang}, B.}, \bibinfo{author}{{Dai}, X.},
  \bibinfo{author}{{Lloyd-Ronning}, N.~M.} \&
  \bibinfo{author}{{M{\'e}sz{\'a}ros}, P.}
\newblock \bibinfo{title}{{Quasi-universal Gaussian Jets: A Unified Picture for
  Gamma-Ray Bursts and X-Ray Flashes}}.
\newblock \emph{\bibinfo{journal}{\apjl}} \textbf{\bibinfo{volume}{601}},
  \bibinfo{pages}{L119--L122} (\bibinfo{year}{2004}).

\bibitem{lamb17}
\bibinfo{author}{{Lamb}, G.~P.} \& \bibinfo{author}{{Kobayashi}, S.}
\newblock \bibinfo{title}{{Electromagnetic counterparts to structured jets from
  gravitational wave detected mergers}}.
\newblock \emph{\bibinfo{journal}{\mnras}} \textbf{\bibinfo{volume}{472}},
  \bibinfo{pages}{4953--4964} (\bibinfo{year}{2017}).

\bibitem{berger14}
\bibinfo{author}{{Berger}, E.}
\newblock \bibinfo{title}{{Short-Duration Gamma-Ray Bursts}}.
\newblock \emph{\bibinfo{journal}{\araa}} \textbf{\bibinfo{volume}{52}},
  \bibinfo{pages}{43--105} (\bibinfo{year}{2014}).

\bibitem{ruiz16}
\bibinfo{author}{{Ruiz}, M.}, \bibinfo{author}{{Lang}, R.~N.},
  \bibinfo{author}{{Paschalidis}, V.} \& \bibinfo{author}{{Shapiro}, S.~L.}
\newblock \bibinfo{title}{{Binary Neutron Star Mergers: A Jet Engine for Short
  Gamma-Ray Bursts}}.
\newblock \emph{\bibinfo{journal}{\apjl}} \textbf{\bibinfo{volume}{824}},
  \bibinfo{pages}{L6} (\bibinfo{year}{2016}).

\bibitem{perego14}
\bibinfo{author}{{Perego}, A.} \emph{et~al.}
\newblock \bibinfo{title}{{Neutrino-driven winds from neutron star merger
  remnants}}.
\newblock \emph{\bibinfo{journal}{\mnras}} \textbf{\bibinfo{volume}{443}},
  \bibinfo{pages}{3134--3156} (\bibinfo{year}{2014}).

\bibitem{bromberg11}
\bibinfo{author}{{Bromberg}, O.}, \bibinfo{author}{{Nakar}, E.},
  \bibinfo{author}{{Piran}, T.} \& \bibinfo{author}{{Sari}, R.}
\newblock \bibinfo{title}{{The Propagation of Relativistic Jets in External
  Media}}.
\newblock \emph{\bibinfo{journal}{\apj}} \textbf{\bibinfo{volume}{740}},
  \bibinfo{pages}{100} (\bibinfo{year}{2011}).

\bibitem{lazzati17}
\bibinfo{author}{{Lazzati}, D.} \emph{et~al.}
\newblock \bibinfo{title}{{Late time afterglow observations reveal a collimated
  relativistic jet in the ejecta of the binary neutron star merger GW170817}}.
\newblock \emph{\bibinfo{journal}{ArXiv e-prints}}  (\bibinfo{year}{2017}).

\bibitem{lamb18}
\bibinfo{author}{{Lamb}, G.~P.}, \bibinfo{author}{{Tanaka}, M.} \&
  \bibinfo{author}{{Kobayashi}, S.}
\newblock \bibinfo{title}{{Transient Survey Rates for Orphan Afterglows from
  Compact Merger Jets}}.
\newblock \emph{\bibinfo{journal}{\mnras}}  (\bibinfo{year}{2018}).

\bibitem{fong12}
\bibinfo{author}{{Fong}, W.} \emph{et~al.}
\newblock \bibinfo{title}{{A Jet Break in the X-Ray Light Curve of Short GRB
  111020A: Implications for Energetics and Rates}}.
\newblock \emph{\bibinfo{journal}{\apj}} \textbf{\bibinfo{volume}{756}},
  \bibinfo{pages}{189} (\bibinfo{year}{2012}).

\bibitem{abbott10}
\bibinfo{author}{Abadie, J.} \emph{et~al.}
\newblock \bibinfo{title}{{Predictions for the Rates of Compact Binary
  Coalescences Observable by Ground-based Gravitational-wave Detectors}}.
\newblock \emph{\bibinfo{journal}{Classical and Quantum Gravity}}
  \textbf{\bibinfo{volume}{27}}, \bibinfo{pages}{173001--+}
  (\bibinfo{year}{2010}).

\bibitem{granot17grb}
\bibinfo{author}{{Granot}, J.}, \bibinfo{author}{{Guetta}, D.} \&
  \bibinfo{author}{{Gill}, R.}
\newblock \bibinfo{title}{{Lessons from the Short GRB 170817A: The First
  Gravitational-wave Detection of a Binary Neutron Star Merger}}.
\newblock \emph{\bibinfo{journal}{\apjl}} \textbf{\bibinfo{volume}{850}},
  \bibinfo{pages}{L24} (\bibinfo{year}{2017}).

\bibitem{lamb17b}
\bibinfo{author}{{Lamb}, G.~P.} \& \bibinfo{author}{{Kobayashi}, S.}
\newblock \bibinfo{title}{{GRB 170817A as a jet counterpart to gravitational
  wave trigger GW 170817}}.
\newblock \emph{\bibinfo{journal}{\mnras}}  (\bibinfo{year}{2018}).

\bibitem{gompertz17}
\bibinfo{author}{{Gompertz}, B.~P.} \emph{et~al.}
\newblock \bibinfo{title}{{The Diversity of Kilonova Emission in Short
  Gamma-Ray Bursts}}.
\newblock \emph{\bibinfo{journal}{ArXiv e-prints}}  (\bibinfo{year}{2017}).

\bibitem{cowperthwaite17}
\bibinfo{author}{{Cowperthwaite}, P.~S.} \emph{et~al.}
\newblock \bibinfo{title}{{The Electromagnetic Counterpart of the Binary
  Neutron Star Merger LIGO/Virgo GW170817. II. UV, Optical, and Near-infrared
  Light Curves and Comparison to Kilonova Models}}.
\newblock \emph{\bibinfo{journal}{\apjl}} \textbf{\bibinfo{volume}{848}},
  \bibinfo{pages}{L17} (\bibinfo{year}{2017}).

\bibitem{kasliwal17}
\bibinfo{author}{{Kasliwal}, M.~M.} \emph{et~al.}
\newblock \bibinfo{title}{{Illuminating gravitational waves: A concordant
  picture of photons from a neutron star merger}}.
\newblock \emph{\bibinfo{journal}{Science}} \textbf{\bibinfo{volume}{358}},
  \bibinfo{pages}{1559--1565} (\bibinfo{year}{2017}).

\bibitem{shappee17}
\bibinfo{author}{{Shappee}, B.~J.} \emph{et~al.}
\newblock \bibinfo{title}{{Early spectra of the gravitational wave source
  GW170817: Evolution of a neutron star merger}}.
\newblock \emph{\bibinfo{journal}{Science}} \textbf{\bibinfo{volume}{358}},
  \bibinfo{pages}{1574--1578} (\bibinfo{year}{2017}).

\bibitem{guillochon17}
\bibinfo{author}{{Guillochon}, J.}, \bibinfo{author}{{Parrent}, J.},
  \bibinfo{author}{{Kelley}, L.~Z.} \& \bibinfo{author}{{Margutti}, R.}
\newblock \bibinfo{title}{{An Open Catalog for Supernova Data}}.
\newblock \emph{\bibinfo{journal}{\apj}} \textbf{\bibinfo{volume}{835}},
  \bibinfo{pages}{64} (\bibinfo{year}{2017}).

\bibitem{margutti17}
\bibinfo{author}{{Margutti}, R.} \emph{et~al.}
\newblock \bibinfo{title}{{The Electromagnetic Counterpart of the Binary
  Neutron Star Merger LIGO/Virgo GW170817. V. Rising X-Ray Emission from an
  Off-axis Jet}}.
\newblock \emph{\bibinfo{journal}{\apjl}} \textbf{\bibinfo{volume}{848}},
  \bibinfo{pages}{L20} (\bibinfo{year}{2017}).

\bibitem{troja17}
\bibinfo{author}{{Troja}, E.} \emph{et~al.}
\newblock \bibinfo{title}{{The X-ray counterpart to the gravitational-wave
  event GW170817}}.
\newblock \emph{\bibinfo{journal}{\nat}} \textbf{\bibinfo{volume}{551}},
  \bibinfo{pages}{71--74} (\bibinfo{year}{2017}).

\bibitem{fruchter02}
\bibinfo{author}{{Fruchter}, A.~S.} \& \bibinfo{author}{{Hook}, R.~N.}
\newblock \bibinfo{title}{{Drizzle: A Method for the Linear Reconstruction of
  Undersampled Images}}.
\newblock \emph{\bibinfo{journal}{\pasp}} \textbf{\bibinfo{volume}{114}},
  \bibinfo{pages}{144--152} (\bibinfo{year}{2002}).

\bibitem{levan17}
\bibinfo{author}{{Levan}, A.~J.} \emph{et~al.}
\newblock \bibinfo{title}{{The Environment of the Binary Neutron Star Merger
  GW170817}}.
\newblock \emph{\bibinfo{journal}{\apjl}} \textbf{\bibinfo{volume}{848}},
  \bibinfo{pages}{L28} (\bibinfo{year}{2017}).

\bibitem{bertin96}
\bibinfo{author}{{Bertin}, E.} \& \bibinfo{author}{{Arnouts}, S.}
\newblock \bibinfo{title}{{SExtractor: Software for source extraction.}}
\newblock \emph{\bibinfo{journal}{\aaps}} \textbf{\bibinfo{volume}{117}},
  \bibinfo{pages}{393--404} (\bibinfo{year}{1996}).

\bibitem{schlafly11}
\bibinfo{author}{{Schlafly}, E.~F.} \& \bibinfo{author}{{Finkbeiner}, D.~P.}
\newblock \bibinfo{title}{{Measuring Reddening with Sloan Digital Sky Survey
  Stellar Spectra and Recalibrating SFD}}.
\newblock \emph{\bibinfo{journal}{\apj}} \textbf{\bibinfo{volume}{737}},
  \bibinfo{pages}{103} (\bibinfo{year}{2011}).

\bibitem{fitzpatrick99}
\bibinfo{author}{{Fitzpatrick}, E.~L.}
\newblock \bibinfo{title}{{Correcting for the Effects of Interstellar
  Extinction}}.
\newblock \emph{\bibinfo{journal}{\pasp}} \textbf{\bibinfo{volume}{111}},
  \bibinfo{pages}{63--75} (\bibinfo{year}{1999}).

\bibitem{alard00}
\bibinfo{author}{{Alard}, C.}
\newblock \bibinfo{title}{{Image subtraction using a space-varying kernel}}.
\newblock \emph{\bibinfo{journal}{\aaps}} \textbf{\bibinfo{volume}{144}},
  \bibinfo{pages}{363--370} (\bibinfo{year}{2000}).

\bibitem{emcee}
\bibinfo{author}{{Foreman-Mackey}, D.}, \bibinfo{author}{{Hogg}, D.~W.},
  \bibinfo{author}{{Lang}, D.} \& \bibinfo{author}{{Goodman}, J.}
\newblock \bibinfo{title}{{emcee: The MCMC Hammer}}.
\newblock \emph{\bibinfo{journal}{\pasp}} \textbf{\bibinfo{volume}{125}},
  \bibinfo{pages}{306} (\bibinfo{year}{2013}).

\bibitem{harris96}
\bibinfo{author}{{Harris}, W.~E.}
\newblock \bibinfo{title}{{A Catalog of Parameters for Globular Clusters in the
  Milky Way}}.
\newblock \emph{\bibinfo{journal}{\aj}} \textbf{\bibinfo{volume}{112}},
  \bibinfo{pages}{1487} (\bibinfo{year}{1996}).

\bibitem{madrid09}
\bibinfo{author}{{Madrid}, J.~P.}, \bibinfo{author}{{Harris}, W.~E.},
  \bibinfo{author}{{Blakeslee}, J.~P.} \& \bibinfo{author}{{G{\'o}mez}, M.}
\newblock \bibinfo{title}{{Structural Parameters of the Messier 87 Globular
  Clusters}}.
\newblock \emph{\bibinfo{journal}{\apj}} \textbf{\bibinfo{volume}{705}},
  \bibinfo{pages}{237--244} (\bibinfo{year}{2009}).

\bibitem{carlson17}
\bibinfo{author}{{Carlson}, N.~L.} \emph{et~al.}
\newblock \bibinfo{title}{{Globular cluster population of the HST frontier
  fields galaxy J07173724+3744224}}.
\newblock \emph{\bibinfo{journal}{ArXiv e-prints}}  (\bibinfo{year}{2017}).

\bibitem{peng09}
\bibinfo{author}{{Peng}, E.~W.} \emph{et~al.}
\newblock \bibinfo{title}{{The Color-Magnitude Relation for Metal-Poor Globular
  Clusters in M87: Confirmation from Deep HST/ACS Imaging}}.
\newblock \emph{\bibinfo{journal}{\apj}} \textbf{\bibinfo{volume}{703}},
  \bibinfo{pages}{42--51} (\bibinfo{year}{2009}).

\bibitem{blanchard17}
\bibinfo{author}{{Blanchard}, P.~K.} \emph{et~al.}
\newblock \bibinfo{title}{{The Electromagnetic Counterpart of the Binary
  Neutron Star Merger LIGO/Virgo GW170817. VII. Properties of the Host Galaxy
  and Constraints on the Merger Timescale}}.
\newblock \emph{\bibinfo{journal}{\apjl}} \textbf{\bibinfo{volume}{848}},
  \bibinfo{pages}{L22} (\bibinfo{year}{2017}).

\bibitem{im17}
\bibinfo{author}{{Im}, M.} \emph{et~al.}
\newblock \bibinfo{title}{{Distance and Properties of NGC 4993 as the Host
  Galaxy of the Gravitational-wave Source GW170817}}.
\newblock \emph{\bibinfo{journal}{\apjl}} \textbf{\bibinfo{volume}{849}},
  \bibinfo{pages}{L16} (\bibinfo{year}{2017}).

\bibitem{davies95}
\bibinfo{author}{{Davies}, M.~B.}
\newblock \bibinfo{title}{{The binary zoo: the calculation of production rates
  of binaries through 2+1 encounters in globular clusters}}.
\newblock \emph{\bibinfo{journal}{\mnras}} \textbf{\bibinfo{volume}{276}},
  \bibinfo{pages}{887--905} (\bibinfo{year}{1995}).

\bibitem{saripirannarayan1998}
\bibinfo{author}{{Sari}, R.}, \bibinfo{author}{{Piran}, T.} \&
  \bibinfo{author}{{Narayan}, R.}
\newblock \bibinfo{title}{{Spectra and Light Curves of Gamma-Ray Burst
  Afterglows}}.
\newblock \emph{\bibinfo{journal}{\apjl}} \textbf{\bibinfo{volume}{497}},
  \bibinfo{pages}{L17--L20} (\bibinfo{year}{1998}).

\bibitem{granotsari02}
\bibinfo{author}{{Granot}, J.} \& \bibinfo{author}{{Sari}, R.}
\newblock \bibinfo{title}{{The Shape of Spectral Breaks in Gamma-Ray Burst
  Afterglows}}.
\newblock \emph{\bibinfo{journal}{\apj}} \textbf{\bibinfo{volume}{568}},
  \bibinfo{pages}{820--829} (\bibinfo{year}{2002}).

\bibitem{kim17}
\bibinfo{author}{{Kim}, S.} \emph{et~al.}
\newblock \bibinfo{title}{{ALMA and GMRT Constraints on the Off-axis Gamma-Ray
  Burst 170817A from the Binary Neutron Star Merger GW170817}}.
\newblock \emph{\bibinfo{journal}{\apjl}} \textbf{\bibinfo{volume}{850}},
  \bibinfo{pages}{L21} (\bibinfo{year}{2017}).

\bibitem{kathirgamaraju18}
\bibinfo{author}{{Kathirgamaraju}, A.}, \bibinfo{author}{{Barniol Duran}, R.}
  \& \bibinfo{author}{{Giannios}, D.}
\newblock \bibinfo{title}{{Off-axis short GRBs from structured jets as
  counterparts to GW events}}.
\newblock \emph{\bibinfo{journal}{\mnras}} \textbf{\bibinfo{volume}{473}},
  \bibinfo{pages}{L121--L125} (\bibinfo{year}{2018}).

\bibitem{gottlieb17}
\bibinfo{author}{{Gottlieb}, O.}, \bibinfo{author}{{Nakar}, E.},
  \bibinfo{author}{{Piran}, T.} \& \bibinfo{author}{{Hotokezaka}, K.}
\newblock \bibinfo{title}{{A cocoon shock breakout as the origin of the $
  \gamma $-ray emission in GW170817}}.
\newblock \emph{\bibinfo{journal}{ArXiv e-prints}}  (\bibinfo{year}{2017}).

\bibitem{metzger10}
\bibinfo{author}{{Metzger}, B.~D.} \emph{et~al.}
\newblock \bibinfo{title}{{Electromagnetic counterparts of compact object
  mergers powered by the radioactive decay of r-process nuclei}}.
\newblock \emph{\bibinfo{journal}{\mnras}} \textbf{\bibinfo{volume}{406}},
  \bibinfo{pages}{2650--2662} (\bibinfo{year}{2010}).

\bibitem{tanaka17}
\bibinfo{author}{{Tanaka}, M.} \emph{et~al.}
\newblock \bibinfo{title}{{Kilonova from post-merger ejecta as an optical and
  near-Infrared counterpart of GW170817}}.
\newblock \emph{\bibinfo{journal}{\pasj}} \textbf{\bibinfo{volume}{69}},
  \bibinfo{pages}{102} (\bibinfo{year}{2017}).

\bibitem{shibata17}
\bibinfo{author}{{Shibata}, M.} \emph{et~al.}
\newblock \bibinfo{title}{{Modeling GW170817 based on numerical relativity and
  its implications}}.
\newblock \emph{\bibinfo{journal}{\prd}} \textbf{\bibinfo{volume}{96}},
  \bibinfo{pages}{123012} (\bibinfo{year}{2017}).

\bibitem{kisaka17}
\bibinfo{author}{{Kisaka}, S.}, \bibinfo{author}{{Ioka}, K.},
  \bibinfo{author}{{Kashiyama}, K.} \& \bibinfo{author}{{Nakamura}, T.}
\newblock \bibinfo{title}{{Scattered Short Gamma-Ray Bursts as Electromagnetic
  Counterparts to Gravitational Waves and Implications of GW170817 and GRB
  170817A}}.
\newblock \emph{\bibinfo{journal}{ArXiv e-prints}}  (\bibinfo{year}{2017}).

\bibitem{fong15}
\bibinfo{author}{{Fong}, W.}, \bibinfo{author}{{Berger}, E.},
  \bibinfo{author}{{Margutti}, R.} \& \bibinfo{author}{{Zauderer}, B.~A.}
\newblock \bibinfo{title}{{A Decade of Short-duration Gamma-Ray Burst Broadband
  Afterglows: Energetics, Circumburst Densities, and Jet Opening Angles}}.
\newblock \emph{\bibinfo{journal}{\apj}} \textbf{\bibinfo{volume}{815}},
  \bibinfo{pages}{102} (\bibinfo{year}{2015}).

\bibitem{mandel17}
\bibinfo{author}{{Mandel}, I.}
\newblock \bibinfo{title}{{The Orbit of GW170817 Was Inclined by Less Than
  28${\deg}$ to the Line of Sight}}.
\newblock \emph{\bibinfo{journal}{\apjl}} \textbf{\bibinfo{volume}{853}},
  \bibinfo{pages}{L12} (\bibinfo{year}{2018}).

\bibitem{granot17a}
\bibinfo{author}{{Granot}, J.}, \bibinfo{author}{{Gill}, R.},
  \bibinfo{author}{{Guetta}, D.} \& \bibinfo{author}{{De Colle}, F.}
\newblock \bibinfo{title}{{Off-Axis Emission of Short GRB Jets from Double
  Neutron Star Mergers and GRB 170817A}}.
\newblock \emph{\bibinfo{journal}{ArXiv e-prints}}  (\bibinfo{year}{2017}).

\bibitem{panaitescu99}
\bibinfo{author}{{Panaitescu}, A.} \& \bibinfo{author}{{M{\'e}sz{\'a}ros}, P.}
\newblock \bibinfo{title}{{Dynamical Evolution, Light Curves, and Spectra of
  Spherical and Collimated Gamma-Ray Burst Remnants}}.
\newblock \emph{\bibinfo{journal}{\apj}} \textbf{\bibinfo{volume}{526}},
  \bibinfo{pages}{707--715} (\bibinfo{year}{1999}).

\bibitem{frailwaxmankulkarni2000}
\bibinfo{author}{{Frail}, D.~A.}, \bibinfo{author}{{Waxman}, E.} \&
  \bibinfo{author}{{Kulkarni}, S.~R.}
\newblock \bibinfo{title}{{A 450 Day Light Curve of the Radio Afterglow of GRB
  970508: Fireball Calorimetry}}.
\newblock \emph{\bibinfo{journal}{\apj}} \textbf{\bibinfo{volume}{537}},
  \bibinfo{pages}{191--204} (\bibinfo{year}{2000}).

\bibitem{sironigiannios2013}
\bibinfo{author}{{Sironi}, L.} \& \bibinfo{author}{{Giannios}, D.}
\newblock \bibinfo{title}{{A Late-time Flattening of Light Curves in Gamma-Ray
  Burst Afterglows}}.
\newblock \emph{\bibinfo{journal}{\apj}} \textbf{\bibinfo{volume}{778}},
  \bibinfo{pages}{107} (\bibinfo{year}{2013}).

\bibitem{lamb2016}
\bibinfo{author}{{Lamb}, G.~P.} \& \bibinfo{author}{{Kobayashi}, S.}
\newblock \bibinfo{title}{{Low-{$\Gamma$} Jets from Compact Stellar Mergers:
  Candidate Electromagnetic Counterparts to Gravitational Wave Sources}}.
\newblock \emph{\bibinfo{journal}{\apj}} \textbf{\bibinfo{volume}{829}},
  \bibinfo{pages}{112} (\bibinfo{year}{2016}).

\bibitem{medvedev2006}
\bibinfo{author}{{Medvedev}, M.~V.}
\newblock \bibinfo{title}{{Electron Acceleration in Relativistic Gamma-Ray
  Burst Shocks}}.
\newblock \emph{\bibinfo{journal}{\apjl}} \textbf{\bibinfo{volume}{651}},
  \bibinfo{pages}{L9--L11} (\bibinfo{year}{2006}).

\bibitem{oppenheimer16}
\bibinfo{author}{{Oppenheimer}, B.~D.} \emph{et~al.}
\newblock \bibinfo{title}{{Bimodality of low-redshift circumgalactic O VI in
  non-equilibrium EAGLE zoom simulations}}.
\newblock \emph{\bibinfo{journal}{\mnras}} \textbf{\bibinfo{volume}{460}},
  \bibinfo{pages}{2157--2179} (\bibinfo{year}{2016}).

\end{thebibliography}

\end{document}